\documentclass[11pt,a4paper]{article}
\usepackage{amsmath, amsfonts, amssymb}
\usepackage{a4wide}
\usepackage{parskip}
\usepackage{enumitem}
\usepackage{xcolor}

\usepackage{hyperref}
\usepackage{booktabs}
\usepackage{caption}
\usepackage{siunitx}
\usepackage{tabularx}
\usepackage{comment}
\usepackage{graphicx}
\usepackage{adjustbox}
\usepackage{multirow}
\usepackage{amsthm}
\usepackage{bm, makecell}
\usepackage{lscape}
\usepackage{threeparttable}
\usepackage{rotating}
\usepackage{floatrow}
\usepackage[noadjust]{cite}

\def\qed{\hfill {$\square$}\goodbreak \medskip}

\newtheorem{theorem}{Theorem}[section]
\newtheorem{lemma}[theorem]{Lemma}
\newtheorem{corollary}[theorem]{Corollary}
\newtheorem{proposition}[theorem]{Proposition}
\newtheorem{definition}[theorem]{Definition}
\newtheorem{example}[theorem]{Example}
\newtheorem{remark}[theorem]{Remark}

\newtheorem{notation}[theorem]{Notation}
\numberwithin{equation}{section}

\newcommand{\lcm}{\textnormal{lcm}}

\usepackage{tikz,xcolor,hyperref}
\usepackage{mathdots}

\definecolor{lime}{HTML}{A6CE39}
\DeclareRobustCommand{\orcidicon}{%
	\begin{tikzpicture}
		\draw[lime, fill=lime] (0,0) 
		circle [radius=0.16] 
		node[white] {{\fontfamily{qag}\selectfont \tiny ID}};
		\draw[white, fill=white] (-0.0625,0.095) 
		circle [radius=0.007];
	\end{tikzpicture}
	\hspace{-2mm}
}

\foreach \x in {A, ..., Z}{%
	\expandafter\xdef\csname orcid\x\endcsname{\noexpand\href{https://orcid.org/\csname orcidauthor\x\endcsname}{\noexpand\orcidicon}}
}

\begin{document}
	\date{}
		\title{$\mathbb{F}_q\mathbb{F}_{q^2}$-additive cyclic codes and their Gray images}

		\author{{\bf Ankit Yadav\footnote{email: {\tt ankityadav10102000@gmail.com}}\orcidA{}  and \bf Ritumoni Sarma\footnote{	email: {\tt ritumoni407@gmail.com}}\orcidC{}} \\ $^{\ast \dagger}$Department of Mathematics\\ Indian Institute of Technology Delhi\\Hauz Khas, New Delhi-110016, India }
  
\maketitle

\begin{abstract}
We investigate additive cyclic codes over the alphabet $\mathbb{F}_{q}\mathbb{F}_{q^2}$, where $q$ is a prime power. First, its generator polynomials and minimal spanning set are determined. Then, examples of $\mathbb{F}_{q^2}$-additive cyclic codes that satisfy the well-known Singleton bound are constructed. Using a Gray map, we produce certain optimal linear codes over $\mathbb{F}_{3}$. Finally, we obtain a few optimal ternary linear complementary dual (LCD) codes from $\mathbb{F}_{3}\mathbb{F}_{9}$-additive codes.

\medskip

			
\medskip
			

\end{abstract}

\section{Introduction}\label{Sec1}
Cyclic codes are a significant family of linear codes because of their extensive algebraic characteristics and practical uses. An advantage of cyclic codes is their structure, which enables efficient encoding and decoding algorithms. Cyclic codes over finite fields were initially introduced by Prange \cite{prange1957cyclic}, and since then, several researchers have considered various rings to study cyclic codes  (c.f. \cite{abualrub2007cyclic, carlet1998Z_2^k, Honold2000linear, markus1999gray,bandi2017negacyclic} and \cite{wolfmann1999negacyclic}). Calderbank and others introduced the notion of $\mathbb{F}_{4}$-additive cyclic codes in \cite{quantum1998} and demonstrated how the class of additive codes outperforms linear codes in minimum Hamming distance. Also, the authors in \cite{quantum1998} constructed binary quantum codes with nice parameters. Huffman studied the $\mathbb{F}_{q^t}$-additive codes in \cite{huffman2013additive} and additive cyclic codes over $\mathbb{F}_{q^t}$ in \cite{huffman2010cyclic} where, for $t=2$, Huffman examined the cyclic codes which are either self-orthogonal or self-dual under trace inner products. Recently, Shi and others investigated the ACD codes over $\mathbb{F}_{4}$ (in \cite{shi2023acd}) and cyclic ACD (in \cite{shi2022cyclicacd}), under two different trace inner products (namely, Euclidean and Hermitian). In \cite{gyan}, the authors generalize the notion of cyclic ACD to $\mathbb{F}_{q^2}$,  for odd prime power $q$. For more on additive codes, one can refer to \cite{gao2021cyclic, rains1999nonbinary, huffman2007additivecyclic} and \cite{sharma2017cyclic}.

Additive codes over the mixed alphabet $\mathbb{Z}_{2}\mathbb{Z}_{4}$ were first introduced by Borges et al. in \cite{borges2010linear}. Rif\`a et al. \cite{rifa}  showed that the perfect $\mathbb{Z}_{2}\mathbb{Z}_{4}$-additive codes are useful in the area of steganography. Abualrub et al. \cite{abualrub2014bbz_} studied $\mathbb{Z}_{2}\mathbb{Z}_{4}$-cyclic codes, and  Shi et al. \cite{shi2019} extended these results to additive codes over  $\mathbb{Z}_{p}\mathbb{Z}_{p^k}$. Later, Aydogdu et al. in \cite{aydogdu2016mathbb} and \cite{aydogdu2017structure} explored additive cyclic codes over $\mathbb{Z}_{2}\mathbb{Z}_{2}[u]$ and $\mathbb{Z}_{2}\mathbb{Z}_{2}[u^3]$ and obtained optimal binary codes. The article \cite{shi_gray} studies Gray images of $\mathbb{Z}_{p^2}$-cyclic codes and $\mathbb{Z}_{p}\mathbb{Z}_{p^2}$-cyclic codes. Lately, numerous researchers have considered various mixed alphabets to study additive codes and derived several linear and quantum codes with good parameters, for instance, \cite{biswas2022quantum, borges2018z, diao2018ℤpℤp, ankit2025acd, sagar2023acd, sagar2024constacyclic,habibul2022,yao2020ℤ} and \cite{yao2020asymptotically}.

A linear code that intersects its dual trivially turns out to be useful in secure communications and data storage; such a code is referred to as an LCD code. The importance of such codes has increased in recent years, particularly in cryptographic applications, due to their efficiency in preventing side-channel and fault injection attacks, as shown in \cite{attacks}.

 In \cite{F2F4}, the authors studied additive cyclic codes over $\mathbb{F}_{2}\mathbb{F}_{4}$ and obtained several examples of optimal binary codes. Motivated by \cite{F2F4}, we in this article extend their work to the alphabet $\mathbb{F}_{q}\mathbb{F}_{q^2}$. Firstly, generator polynomials and a minimal spanning set of an $\mathbb{F}_{q}\mathbb{F}_{q^2}$-additive cyclic code are determined. We construct many additive codes over $\mathbb{F}_{q^2}$ of various lengths, which are optimal with respect to the Singleton bound. Several ternary optimal codes are derived from $\mathbb{F}_{3}\mathbb{F}_{9}$-additive cyclic codes as Gray images. Furthermore, we construct LCD codes over $\mathbb{F}_{3}$ from $\mathbb{F}_{3}\mathbb{F}_{9}$-additive codes.

 In the forthcoming section, the necessary preliminaries for $\mathbb{F}_{q}\mathbb{F}_{q^2}$-additive cyclic codes are presented.  We split Section \ref{Sec3} into three subsections. In the first, generator polynomials are computed; in the second, a minimal spanning set is determined, and in the third, the dual code is studied. Ternary optimal codes with good parameters are obtained with the help of a Gray map in Section \ref{Sec4}. We present a construction of a $q$-ary LCD code from a $\mathbb{F}_{q}\mathbb{F}_{q^2}$-additive code and construct several ternary optimal LCD codes in Section \ref{Sec5}. We end the article with a short discussion in Section \ref{Sec6}.

\section{Preliminaries} \label{Sec2}
Throughout the article, $\mathbb{F}_{q}$ is a finite field, where $q$ is its cardinality. Recall $\mathbb{F}_{q^2} \cong  \frac{\mathbb{F}_q[x]}{\langle f(x)\rangle}$ for an irreducible quadratic polynomial $f(x)$ over $\mathbb{F}_{q}$. Let $\omega$ be a zero of $f(x)$. Then every element $z \in \mathbb{F}_{q^2}$ has the form $z = b+\omega c$ for $b,c \in \mathbb{F}_{q}$. \par
\begin{definition}
    A subspace $C$ of the $\mathbb{F}_{q}$-vector space $\mathbb{F}_{q^2}^{n}$ is called an  $\mathbb{F}_{q^2}$-additive code. The parameter $n$ is the length of $C$.
\end{definition}
\begin{definition}
    A submodule $C$ of the $\mathbb{F}_{q}[x]$-module $\frac{\mathbb{F}_{q^2}[x]}{\langle x^n-1 \rangle}$ is called an $\mathbb{F}_{q^2}$-additive cyclic code.
\end{definition}

For $\alpha, \beta \in \mathbb{N}$, an element of the $\mathbb{F}_{q}$-vector space $\mathbb{F}_{q}^{\alpha} \times \mathbb{F}_{q^2}^{\beta}$ is denoted by $( u_{0}, u_{1}, \ldots, u_{\alpha-1}\mid u_{0}', u_{1}', \ldots, u_{\beta-1}')$, where $u_{i} \in \mathbb{F}_{q}$ and $u_{j}' \in \mathbb{F}_{q^2}$.
\begin{definition}
 An $\mathbb{F}_{q}$-subspace $\mathcal{C}$ of $\mathbb{F}_{q}^{\alpha} \times \mathbb{F}_{q^2}^{\beta}$ is referred to as an $\mathbb{F}_{q}\mathbb{F}_{q^2}$-additive code and $(\alpha, \beta)$ is the block length of $\mathcal{C}$. 
\end{definition}
    Throughout the discussion, we assume without mentioning every time the block length of the $\mathbb{F}_{q}\mathbb{F}_{q^2}$-additive code to be $(\alpha,\beta)$.
\begin{remark}
    Observe that $\mathbb{F}_{q}^{\alpha} \times \mathbb{F}_{q^2}^{\alpha} \cong (\mathbb{F}_{q}\mathbb{F}_{q^2})^{\alpha}$. An $\mathbb{F}_{q}\mathbb{F}_{q^2}$-additive code with block length $(\alpha, \alpha)$ is called an $\mathbb{F}_{q}\mathbb{F}_{q^2}$-additive code of length $\alpha$.
\end{remark}
\begin{definition}
For $\mathbf{u}, \mathbf{v} \in \mathbb{F}_{q}^{\alpha} \times \mathbb{F}_{q^2}^{\beta}$, we shall consider throughout the inner product given by
    \begin{equation} \label{innerprod}
            \langle \mathbf{u}, \mathbf{v} \rangle := \omega \sum\limits_{i=0}^{\alpha-1} u_{i}v_{i} + \sum\limits_{j=0}^{\beta-1} u_{j}'v_{j}'  \in \mathbb{F}_{q^2}.
    \end{equation}
\end{definition}
 Obviously, in this case, the dual code is also an $\mathbb{F}_{q}\mathbb{F}_{q^2}$-additive code.

\begin{notation}
    For an  $\mathbb{F}_{q}\mathbb{F}_{q^2}$-additive code $\mathcal{C}$, the image of $C$ under the projection map onto the first $\alpha$ components and onto the last $\beta$ components are respectively denoted by $\mathcal{C}_{\alpha}$ and $\mathcal{C}_{\beta}.$ 
\end{notation}
\begin{definition}
    Any $(n, q^k, d)_{q}$-code must satisfy the inequality $q^k \leq q^{n-d+1}$, which is the Singleton bound (for $k$). Any linear code satisfying this bound is called an MDS (that abbreviates maximum distance separable) code.
\end{definition}
Set $\bar{\mathbf{u}} = (\mathbf{u} \mid \mathbf{u}')$, where $\mathbf{u} = (u_{0},u_{1}, \ldots, u_{\alpha-1}) \in \mathbb{F}_{q}^{\alpha}$ and $\mathbf{u}'= (u_{0}', u_{1}', \ldots, u_{\beta -1}') \in \mathbb{F}_{q^2}^{\beta}$. Then, for $i \in \mathbb{N}$, $\bar{\mathbf{u}}^{(i)}$ stands for the word obtained from $\bar{\mathbf{u}}$ by applying the right cyclic shift $i$-times. For example, $\bar{\mathbf{u}}^{(1)} = (u_{\alpha-1},u_{0}, \ldots, u_{\alpha-2} \mid u_{\beta-1}', u_{0}', \ldots, u_{\beta -2}')$.

\begin{definition}
    An $\mathbb{F}_{q}\mathbb{F}_{q^2}$-additive code $\mathcal{C}$ is cyclic if  $\mathcal{C}$ is closed under the right cyclic shift, that is, $\bar{\mathbf{u}}^{(1)} \in \mathcal{C} $ whenever $\bar{\mathbf{u}} \in \mathcal{C}.$
\end{definition}

\begin{notation}
     If $\mathbf{a} = (a_{0},a_{1},\ldots,a_{r-1})$, then write $\mathbf{a}(x) = a_{0}+a_{1}x+\cdots+a_{r-1}x^{r-1}$.
\end{notation}
Denote the product ring $\frac{\mathbb{F}_{q}[x]}{\langle x^\alpha -1 \rangle} \times \frac{\mathbb{F}_{q^2}[x]}{\langle x^\beta -1 \rangle}$ by $\mathcal{R}_{\alpha,\beta}$. There is a bijection between $\mathbb{F}_{q}^{\alpha} \times \mathbb{F}_{q^2}^{\beta}$ and $\mathcal{R}_{\alpha, \beta}$ given by 
\begin{equation*}
    (\mathbf{u}\mid \mathbf{u}') \mapsto (\mathbf{u}(x) \mid \mathbf{u}'(x)). 
\end{equation*}
For any $s(x) = \sum\limits_i s_{i} x^{i} \in \mathbb{F}_{q}[x]$ and $(\mathbf{u}(x)\mid \mathbf{u}'(x)) \in \mathcal{R}_{\alpha,\beta}$, define the multiplication  $* : \mathbb{F}_{q}[x] \times \mathcal{R}_{\alpha,\beta} \rightarrow \mathcal{R}_{\alpha,\beta}$ by
\begin{equation*}
    s(x)*(\mathbf{u}(x) \mid \mathbf{u}'(x)) = (s(x)\mathbf{u}(x) \mid s(x)\mathbf{u}'(x)).
\end{equation*}  
The above multiplication $*$ is well-defined and $\mathcal{R}_{\alpha,\beta}$ is an $\mathbb{F}_{q}[x]$-module under the scalar multiplication $*$.
\begin{theorem}
    An $\mathbb{F}_{q}\mathbb{F}_{q^2}$-additive code $\mathcal{C}$ is cyclic if and only if $\mathcal{C}$ is an $\mathbb{F}_{q}[x]$-submodule of $\mathcal{R}_{\alpha,\beta}$.
\end{theorem}
\begin{proof}
    Let $\bar{\mathbf{u}}=(\mathbf{u}\mid \mathbf{u}') \in \mathbb{F}_{q}^{\alpha} \times \mathbb{F}_{q^2}^{\beta} $ and let $\bar{\mathbf{u}}(x)=(\mathbf{u}(x) \mid \mathbf{u}'(x)) \in \mathcal{R}_{\alpha,\beta}$ be the corresponding polynomial in $\mathcal{R}_{\alpha,\beta}$. Observe $x * \bar{\mathbf{u}}(x) = \bar{\mathbf{u}}^{(1)}(x)$. Thus, the theorem follows. \qed
\end{proof}
Now, we prove a couple of elementary lemmas.

\begin{lemma} \label{keylemma}
    Suppose $A$ is a commutative ring with unity. If $\eta: V\rightarrow W$ is an $A$-module homomorphism such that $\eta(V) = Aw$, then $V = Ker(\eta)+Av$, for any $v \in \eta^{-1}(w)$.
\end{lemma}
\begin{proof}
    It follows directly from the First Isomorphism Theorem for modules. \qed
\end{proof}

\begin{lemma} \label{keylemma1}
    Suppose $A$ is a commutative ring with unity. Then every $A$-submodule of $A/I$ is an ideal of the quotient ring $A/I$. 
\end{lemma}
\begin{proof}
    Suppose $V$ is an $A$-submodule of $A/I$. Let $c+I = c'+I$ for $c,c' \in A$ so that $c-c' \in I$. Then $cm-c'm = (c-c')m = 0$ in $A/I$ for $m \in V$. So, $V$ is an $A/I$-submodule of $A/I$ with $(c+I)m = cm$. \qed
\end{proof}
\section{The additive cyclic code over $\mathbb{F}_{q}\mathbb{F}_{q^2}$} \label{Sec3}
\subsection{Generators }
First, we give a description of polynomials that generate the additive cyclic code over $\mathbb{F}_{q^2}$.
\begin{theorem} \label{generators_F_q^2}
     If $n$ is the length of the $\mathbb{F}_{q^2}$-additive cyclic code $C$, then $$C=\langle g(x)+\omega h(x), \omega k(x) \rangle,$$ where $g(x), k(x)$ and $ h(x) \in \mathbb{F}_{q}[x]$ with $g(x)$, $k(x)$ $\mid x^n-1$ in $\mathbb{F}_{q}[x]$.
\end{theorem}
\begin{proof}
Consider the $\mathbb{F}_{q}[x]$-module homomorphism 
$\psi : C  \rightarrow \frac{\mathbb{F}_{q}[x]}{\langle x^n-1\rangle}$ given by 
\begin{equation}\label{psimap}
    \psi\left(c_{1}(x)+\omega c_{2}(x)+\langle x^n-1\rangle\right) = c_{1}(x)+\langle x^n-1\rangle,
\end{equation}
where $c_{1}(x), c_{2}(x) \in \mathbb{F}_{q}[x]$. By Lemma \ref{keylemma1}, $\psi(C)$ is an ideal of $\frac{\mathbb{F}_{q}[x]}{\langle x^n-1\rangle}$. Therefore, by Theorem 7.2.3 of \cite{lingxing}, $\psi(C) =\langle g(x) \rangle$, where $g(x) \mid x^n-1$. Moreover, Ker($\psi$) = $\langle \omega k(x) \rangle$, where $k(x) \in \mathbb{F}_{q}[x]$ and $k(x) \mid$ $x^n-1$. Suppose $g(x)+\omega h(x)$ is a pre-image of $g(x)$ so that $h(x) \in \mathbb{F}_{q}[x]$. Therefore, by Lemma \ref{keylemma}, it follows that $C=\langle g(x)+\omega h(x), \omega k(x) \rangle$. \qed
\end{proof}
Next, we give a description of polynomials that generate the additive cyclic code over the mixed alphabet.
\begin{theorem} \label{generators}
    Suppose $\mathcal{C}$ is an $\mathbb{F}_{q}\mathbb{F}_{q^2}$-additive cyclic code. Then $$\mathcal{C} = \langle (s(x) \mid l(x)),(0\mid \omega k(x)), (0 \mid g(x)+\omega h(x))  \rangle,$$
    where $s(x), g(x), h(x)$ and $k(x) \in \mathbb{F}_{q}[x]$ with $s(x) \mid x^\alpha -1$ and $g(x), k(x) \mid x^\beta -1$ and  $l(x) \in \mathbb{F}_{q^2}[x]$. Further, $k(x)$ divides $ h(x) \frac{x^\beta -1}{g(x)}$ and $\frac{x^{\alpha}-1}{s(x)}l(x) \in \langle  g(x)+\omega h(x), \omega k(x) \rangle$.
\end{theorem}
\begin{proof}
Consider the $\mathbb{F}_{q}[x]$-module homomorphism $\pi_{1}: \mathcal{C} \rightarrow \frac{\mathbb{F}_{q}[x]}{\langle x^{\alpha}-1\rangle}$ defined by $(\mathbf{u}(x)\mid \mathbf{u}'(x)) \mapsto \mathbf{u}(x)$. Then, by Lemma \ref{keylemma1}, $\pi_{1}(\mathcal{C})$ is an ideal of $\frac{\mathbb{F}_{q}[x]}{\langle x^{\alpha}-1 \rangle}$. Therefore, for some $s(x) \in \mathbb{F}_{q}[x]$ and $s(x) \mid x^{\alpha} -1$, we have  $\pi_{1}(\mathcal{C}) = \langle s(x) \rangle$. Since $s(x) \in \pi_{1}(\mathcal{C})$, there exists $l(x) \in \frac{\mathbb{F}_{q^2}[x]}{\langle x^{\beta}-1\rangle}$ such that $(s(x)\mid l(x)) \in \mathcal{C} $. The Kernel of the map $\pi_{1}$ is given by $$K = Ker(\pi_{1}) = \left\{(0\mid \mathbf{u}'(x)) \in \mathcal{C} \mid \mathbf{u}'(x) \in \frac{\mathbb{F}_{q^2}[x]}{\langle x^{\beta}-1 \rangle} \right\}.$$ Consider the set $N = \left\{ \mathbf{u}'(x) : (0\mid \mathbf{u}'(x)) \in \mathcal{C} \right\} \subseteq \frac{\mathbb{F}_{q^2}[x]}{\langle 
 x^{\beta}-1\rangle}$. Since $\mathcal{C}$ is an $\mathbb{F}_{q}[x]$-submodule of $\mathcal{R}_{\alpha,\beta}$, $N$ is an additive cyclic code over $\mathbb{F}_{q^2}$. By Theorem \ref{generators_F_q^2}, $N = \langle g(x)+\omega h(x), \omega k(x)  \rangle$, where $g(x),h(x), k(x) \in \mathbb{F}_{q}[x]$ and $g(x), k(x) \mid $ $x^\beta -1$ in $\mathbb{F}_{q}[x]$. Therefore, $K = \langle (0\mid g(x)+\omega h(x)), (0\mid \omega k(x)) \rangle$.
Then, it follows from Lemma \ref{keylemma} that $\mathcal{C} = \langle (s(x) \mid l(x)),(0\mid \omega k(x)), (0 \mid g(x)+\omega h(x))  \rangle$.\par
Moreover, $\frac{x^{\alpha}-1}{s(x)}*(s(x)\mid l(x)) = (0\mid \frac{x^{\alpha}-1}{s(x)} l(x)) \in \mathcal{C}$, which implies that the polynomial $ \frac{x^{\alpha}-1}{s(x)} l(x) $ belongs to the $\mathbb{F}_{q}[x]$-submodule $ \langle g(x)+\omega h(x), \omega k(x) \rangle$. \qed
\end{proof}

\subsection{Minimal Spanning set}
Now, we determine an $\mathbb{F}_{q}$-basis of the code.
\begin{proposition}\label{fq_basis}
    Suppose $C $ is an $\mathbb{F}_{q^2}$-additive cyclic code given by $C=\langle g(x)+\omega h(x), \omega k(x) \rangle$, where $g(x),h(x)$ and $k(x) \in \mathbb{F}_{q}[x]$ with $g(x)$ and $k(x)$ dividing $x^n-1$, then the set $$ T =\left\{ x^{i}(g(x)+\omega h(x)) : 0 \leq i\leq n-\deg(g)-1 \right\} \cup \left\{\omega x^{j}k(x) : 0\leq j \leq n-\deg(k)-1  \right\}$$ is an $\mathbb{F}_{q}$-basis of $C$. 
\end{proposition}
\begin{proof}
    Observe that the set $T$ is linearly independent over $\mathbb{F}_{q}$. Since $T \subseteq C$, $\text{Span}_{\mathbb{F}_{q}}(T) \subseteq C$. It is enough to show that $C \subseteq \text{Span}_{\mathbb{F}_{q}}(T)$. Suppose $\mathbf{c}(x)$ is a codeword in polynomial form. Then, we have $\mathbf{c}(x) = e_{1}(x)(g(x)+\omega h(x))+\omega e_{2}(x)k(x)$ for $e_{1}(x),e_{2}(x) \in \mathbb{F}_{q}[x]
    $. Since $\omega e_{2}(x)k(x) \in \text{Ker}(\psi) = \langle \omega k(x) \rangle$, we have $\omega e_{2}(x)k(x) \equiv \omega\Tilde{e}_{2}(x)k(x)$ in $\frac{\mathbb{F}_{q^2}[x]}{\langle x^n-1 \rangle}$, where $\Tilde{e}_{2}(x)\in \mathbb{F}_{q}[x]$ with $\deg(\Tilde{e}_{2}(x)) < n-\deg(k)$. 
    Also, 
    \begin{eqnarray*}
        \psi(e_{1}(x)g(x)+\omega e_{1}(x)h(x)) &=& e_{1}(x)g(x) \\
    &\equiv& \Tilde{e}_{1}(x)g(x) \;\; \text{in} \;\; \frac{\mathbb{F}_{q}[x]}{\langle x^n-1 \rangle},   
    \end{eqnarray*}
    for some $\Tilde{e}_{1}(x) \in \mathbb{F}_{q}[x]$ with $\deg(\Tilde{e}_{1}(x)) < n-\deg(g).$ Then $$\psi\left(e_{1}(x)(g(x)+\omega h(x))  - \Tilde{e}_{1}(x)(g(x)+\omega h(x))\right) = 0.$$ This implies $e_{1}(x)(g(x)+\omega h(x)) - \Tilde{e}_{1}(x)(g(x)+\omega h(x)) \in \text{Ker}(\psi) = \langle \omega k(x) \rangle$. Thus, there exists $e'(x) \in \mathbb{F}_{q}[x]$ with $\deg(e'(x)) < n- \deg(k)$ such that $e_{1}(x)(g(x)+\omega h(x)) = \Tilde{e}_{1}(x)(g(x)+\omega h(x)) + \omega e'(x)k(x)$. Therefore, $\mathbf{c}(x) \equiv \Tilde{e}_{1}(x)(g(x)+\omega h(x))+ \omega(\Tilde{e}_{2}(x)+e'(x))k(x)$  so that $\mathbf{c}(x)$ is an $\mathbb{F}_{q}$-span of $T$.
\end{proof}
For an additive cyclic code over $\mathbb{F}_{q}\mathbb{F}_{q^2},$ the following result presents its minimal spanning subset.
\begin{theorem} \label{minimalspanningset}
    Suppose $\mathcal{C}$ denotes a code which is described as in Theorem \ref{generators}. Then, a minimal spanning set of $\mathcal{C}$ is given by $S = \bigcup\limits_{i=1}^{3} S_{i}$, where 
    \begin{eqnarray*}
         S_{1} &=& \bigcup\limits_{i=0}^{\alpha-\deg(s)-1}x^{i}*(s(x)\mid l(x)),\\
          S_{2} &=& \bigcup\limits_{i=0}^{\beta-\deg(g)-1}x^{i}*(0 \mid g(x)+\omega h(x)),\\
           S_{3} &=& \bigcup\limits_{i=0}^{\beta-\deg(k)-1}x^{i}*(0 \mid \omega k(x)).
    \end{eqnarray*}
\end{theorem}
\begin{proof}
Let $\mathbf{c}(x) \in \mathcal{C}$. Then there exists $t_{1}(x), t_{2}(x), t_{3}(x) \in \mathbb{F}_{q}[x] $ such that $\mathbf{c}(x) = t_{1}(x)*(0\mid g(x)+\omega h(x)) + t_{2}(x)*(0\mid \omega k(x)) + t_{3}(x)*(s(x)\mid l(x))$. By Proposition \ref{fq_basis}, we can assume that $\deg(t_{1}) < \beta -\deg(g)$ and $\deg(t_{2}) < \beta -\deg(h)$. If $\deg(t_{3}) < \alpha -\deg(s)$, then $t_{3}(x)*(s(x)\mid l(x)) \in $ Span($S_{3}$) and hence $\mathbf{c}(x) \in $ Span($S$). If not, by division algorithm, there exists $q_{1}(x), r_{1}(x) \in \mathbb{F}_{q}[x]$ such that 
    $$t_{3}(x) = \frac{x^{\alpha}-1}{s(x)}q_{1}(x)+r_{1}(x),$$
    where $r_{1} = 0$ or $\deg(r_{1}) < \alpha - \deg(s)$. Then 
    \begin{eqnarray*}
     t_{3}(x)*(s(x)\mid l(x)) &=& \left(r_{1}(x)s(x)\mid \frac{x^{\alpha}-1}{s(x)}q_{1}(x)l(x)+r_{1}(x)l(x)\right) \\
     &=& r_{1}(x)*\left(s(x)\mid l(x)\right) + \left(0 \mid \frac{x^{\alpha}-1}{s(x)}q_{1}(x)l(x)\right).
    \end{eqnarray*}
    Since $(s(x)\mid l(x)) \in \mathcal{C}, \left(0 \mid \frac{x^{\alpha}-1}{s(x)}q_{1}(x)l(x)\right) \in \mathcal{C}$. By Theorem \ref{generators}, $\frac{x^{\alpha}-1}{s(x)}q_{1}(x)l(x) \in K = \langle g(x)+\omega h(x), \omega k(x) \rangle$ and therefore there exists $t_{4}(x), t_{5}(x) \in \mathbb{F}_{q}[x]$ with $\deg(t_{4}) < \beta -\deg(g)$ and $\deg(t_{5}) < \beta-\deg(k)$ such that $\frac{x^{\alpha}-1}{s(x)}q_{1}(x)l(x) = t_{4}(x)(g(x)+\omega h(x))+t_{5}(x)(\omega k(x))$. Hence 
    \begin{eqnarray*}
        \mathbf{c}(x) &=& t_{1}(x)*\left(0 \mid g(x)+ \omega h(x)\right)+t_{2}(x)*(0\mid \omega k(x))+t_{3}(x)*(s(x)\mid l(x))\\ 
        &=&  t_{1}(x)*(0\mid g(x)+\omega h(x))+t_{2}(x)*(0\mid \omega k(x))+ r_{1}(x)*(s(x)\mid l(x)) +\\
        &&t_{4}(x)*(0\mid g(x)+\omega h(x))+t_{5}(x)(0\mid \omega k(x)) \\
        &=& \left(t_{1}(x)+t_{4}(x))*(0\mid g(x)+\omega h(x)\right)+ (t_{2}(x)+t_{5}(x))*(0\mid \omega k(x))+\\
        &&r_{1}(x)*(s(x)\mid l(x)).
    \end{eqnarray*}
This implies $\mathbf{c}(x) \in$ Span($S$) and therefore $\mathcal{C}= $Span($S$). \qed
\end{proof}
\begin{corollary}
    Suppose $\mathcal{C} = \langle (s(x) \mid l(x)),(0\mid \omega k(x)), (0 \mid g(x)+\omega h(x))  \rangle $ be an $\mathbb{F}_{q}\mathbb{F}_{q^2}$-additive cyclic code. Then $|\mathcal{C}| = q^{\alpha - \deg(s)} q^{\beta -\deg(g)} q^{\beta -\deg(k)}$.
\end{corollary}
\subsection{The Dual}
\begin{proposition}\label{dual}
The dual $\mathcal{C}^{\perp}$ of an $\mathbb{F}_{q}\mathbb{F}_{q^2}$-additive cyclic code $\mathcal{C}$ is also an additive cyclic code over $\mathbb{F}_{q}\mathbb{F}_{q^2}$.
\end{proposition}
\begin{proof}
   It is enough to show that $\langle \bar{\mathbf{u}}, \bar{\mathbf{v}}^{(1)} \rangle = 0$ for $\bar{\mathbf{u}} \in \mathcal{C}$ and $\bar{\mathbf{v}} \in \mathcal{C}^{\perp}$.
   Let $\gamma = \lcm(\alpha, \beta)$. Since $\mathcal{C}$ is additive cyclic, $\bar{\mathbf{u}}^{(\gamma-1)} \in \mathcal{C}$. Then 
    \begin{eqnarray*}
        0 &=& \langle \bar{\mathbf{u}}^{(\gamma-1)}, \bar{\mathbf{v}} \rangle \\
        &=& \omega (u_{1}v_{0} + u_{2}v_{1} + \cdots + u_{0}v_{\alpha-1}) + (u_{1}'v_{0}' + u_{2}'v_{1}' + \cdots + u_{0}'v_{\beta-1}') \\
        &=& \langle \bar{\mathbf{u}}, \bar{\mathbf{v}}^{(1)} \rangle. 
    \end{eqnarray*}
    \qed
\end{proof}
The following result is a consequence of Proposition \ref{dual} and Theorem \ref{generators}.
\begin{theorem}
   The dual $\mathcal{C}^\perp$ of an additive cyclic code $\mathcal{C}$ over $\mathbb{F}_{q}\mathbb{F}_{q^2}$ is given by $$\mathcal{C}^{\perp} = \langle  (s'(x) \mid l'(x)),(0\mid \omega k'(x)), (0 \mid g'(x)+\omega h'(x))  \rangle,$$
    where $s'(x), g'(x), h'(x)$ and $k'(x) \in \mathbb{F}_{q}[x]$ with $s'(x) \mid x^\alpha -1$ and $g'(x), k'(x) \mid x^\beta -1$ and $l'(x) \in \mathbb{F}_{q^2}[x]$.
\end{theorem}
\begin{proof}
    It is a consequence of Theorem \ref{generators}. \qed
\end{proof}
Using Theorem \ref{generators_F_q^2}, we construct some optimal $\mathbb{F}_{q^2}$-additive cyclic codes that attain the Singleton bound. These codes are listed in Table \ref{table1}. In Table \ref{table1}, $u$ denotes the root of the defining polynomial $x^2+x+1$ (for $\mathbb{F}_{4}$) or $x^3+x+1$ (for $\mathbb{F}_{8}$) in $\mathbb{F}_{2}[x]$.
\begin{table}
     \centering
     
    \begin{tabular}{|l|l|l|p{5cm}|p{5cm}|l|}
    \hline
      \multirow{2}{*}{$q$} & \multirow{2}{*}{$n$} & \multicolumn{3}{|c|}{Generators}  & \multirow{2}{*}{Parameters}\\
\cline{3-5}
& & $g(x)$ & $h(x)$ & $k(x)$ & \\
        \Xhline{3\arrayrulewidth}
   $4$ & $5$ & $1$ & $ x^2 + ux$ & $x^4 + x^3 + x^2 + x + 1$ & $(5,(4^2)^3, 3)$ \\
   \hline
  $4$ & $6$ & $x^2+u$ & $ x^4 + x^3 + ux+u^2$ & $x^6 + 1$ & $(6,(4^2)^2, 5)$ \\
   \hline
   $4$ & $7$ & $1$ & $ x^3 + ux^2 + x$ & $ x^6 + x^5 + x^4 + x^3 + x^2 + x + 1$ & $(7,(4^2)^4, 4)$ \\
   \hline
   $4$ & $8$ & $1$ & $ x^3 + x^2+ ux$ & $x^6 + x^4 + x^2 + 1$ & $(8,(4^2)^5, 4)$ \\
   \hline
   $4$ & $9$ & $1$ & $ x^5 + x^4+x^3+ux$ & $x^8 + ux^7 + u^2x^6 + x^5 + ux^4+u^2x^3+x^2+ux+u^2 $ & $(9,(4^2)^5, 5)$ \\
   \hline
   $4$ & $10$ & $1$ & $x^7 + x^6 + x^5 + ux^3 + x^2+u^2x$ & $ x^{10} + 1$ & $(10,(4^2)^5, 6)$ \\
   \hline
   $4$ & $13$ & $1$ & $x^5 + x^3 + ux^2+u^2x$ & $x^6+ux^5+u^2x^3+ux+1$ & $(13,(4^2)^{10}, 4)$ \\
   \hline
   $4$ & $15$ & $1$ & $x^2 + x$ & $x^4+x+u^2$ & $(15,(4^2)^{13}, 3)$ \\
   \hline
   $4$ & $17$ & $1$ & $x^7 + x^6 + ux^3 + u^2x$ & $x^8 + ux^7 + ux^5 + ux^4 + ux^3 + ux + 1$ & $(17,(4^2)^{13}, 5)$ \\
   \hline
   $8$ & $5$ & $1$ & $x^2+ux$ & $x^4 + x^3 + x^2 + x + 1$ & $(5,(8^2)^{3}, 3)$ \\
   \hline
   $8$ & $6$ & $1$ & $x^3+x^2+ux$ & $x^6+ 1$ & $(6,(8^2)^{3}, 4)$ \\
   \hline
   $8$ & $7$ & $1$ & $x^2+x$ & $x^4 + u^5x^3 + u^4x^2 + x + u^4$ & $(7,(8^2)^{5}, 3)$ \\
   \hline
   $8$ & $8$ & $1$ & $x^4+x^3+ux^2+u^3x$ & $x^8+ 1$ & $(8,(8^2)^{4}, 5)$ \\
   \hline
   $8$ & $9$ & $1$ & $x^3+x^2+ux$ & $x^6+u^6x^5+ux^4+u^5x^3+ux^2+u^6x+1$ & $(9,(8^2)^{6}, 4)$ \\
   \hline
   $8$ & $10$ & $1$ & $x^5+x^4+ux^3+u^6x^2+u^2x$ & $x^{10}+1$ & $(10,(8^2)^{5}, 6)$ \\
   \hline
   $8$ & $11$ & $1$ & $x^6+x^4+ux^3+u^5x^2+x$ & $x^{10}+x^9+x^8+x^7+x^6+x^5+x^4+x^3+x^2+x+1$ & $(11,(8^2)^{6}, 6)$ \\
   \hline
   $8$ & $13$ & $1$ & $x^2+x$ & $x^{4}+u^6x^3+u^3x^2+u^6x+1$ & $(13,(8^2)^{11}, 3)$ \\
   \hline
   $8$ & $15$ & $1$ & $x^2+ux$ & $x^{4}+x^3+1$ & $(15,(8^2)^{13}, 3)$ \\
   \hline 
   $8$ & $17$ & $1$ & $x^6+ux^5+u^3x^3+ux^2+u^3x$ & $x^8+x^5+x^4+x^3+1$ & $(17,(8^2)^{13}, 5)$ \\
   \hline 
   
   \hline
    \end{tabular}
    \caption{$\mathbb{F}_{q^2}$-additive cyclic codes}
    \label{table1}
\end{table}

\section{Gray image of $\mathbb{F}_{q}\mathbb{F}_{q^2}$-additive cyclic codes} \label{Sec4}
 Any element $z \in \mathbb{F}_{q^2}$ can be written as $z = b+\omega c$, where $b,c \in \mathbb{F}_{q}$. Consider the Gray map $\phi : \mathbb{F}_{q^2}^{\beta} \rightarrow \mathbb{F}_{q}^{2\beta}$ given by
 $$\phi( b_{0}+\omega c_{0},b_{1}+\omega c_{1},\ldots,b_{\beta-1}+\omega c_{\beta-1}) = (b_{0}+c_{0},b_{1}+c_{1},\ldots,b_{\beta-1}+c_{\beta-1},c_{0},c_{1},\ldots,c_{\beta-1}).$$
 This map gets extended to 
 $\Phi : \mathbb{F}_{q}^{\alpha}\mathbb{F}_{q^2}^{\beta} \rightarrow \mathbb{F}_{q}^{\alpha+2\beta}$ 
 \begin{equation} \label{phi}
      \Phi(\mathbf{u}\mid \mathbf{u}') 
    = (\mathbf{u}, \phi(\mathbf{u}')).
 \end{equation}

 \begin{lemma}
     The map $\Phi$ defined in Eq. (\ref{phi}) is bijective and linear over $\mathbb{F}_{q}$.
 \end{lemma}
 \begin{lemma}
     Suppose $\mathcal{C}$ is an $\mathbb{F}_{q}\mathbb{F}_{q^2}$-additive code. Then the minimum distance of $\Phi(\mathcal{C})$ is not less than the minimum distance of $\mathcal{C}$.
 \end{lemma}
 \begin{proof}
    See Lemma 3 of \cite{F2F4}.
 \end{proof}
The next theorem characterizes the Gray image.
\begin{theorem}
    If $\mathcal{C}$ is an additive cyclic code over $\mathbb{F}_{q}\mathbb{F}_{q^2}$ and $\Phi$ is as in Eq. (\ref{phi}), then $\Phi(\mathcal{C})$ is
    \begin{enumerate}
        \item a quasi-cyclic code over $\mathbb{F}_{q}$ of length $\alpha+2\beta$ and index $3$ if $\alpha = \beta$,
        \item a generalised quasi-cyclic code over $\mathbb{F}_{q}$ having block length $(\alpha,2\beta)$ if $\alpha\neq \beta$ and $\gcd(\alpha+2\beta,3) = 3$, and
        \item equivalent to a cyclic code over $\mathbb{F}_{q}$ of length $\alpha+2\beta$ if $\alpha\neq \beta$ and $\gcd(\alpha+2\beta,3) = 1$.
    \end{enumerate}
\end{theorem}
\begin{proof}
    It is similar to Theorem 14 of \cite{F2F4}.
\end{proof}

Using Theorem \ref{generators} and the Gray map $\Phi$ defined in Eq. \eqref{phi}, optimal ternary linear codes are obtained. Table \ref{table2} presents these codes, where $\omega$ is a zero of the polynomial $x^2+1 \in \mathbb{F}_{3}[x]$. With the help of the database \cite{codetable}, we verify the optimality of these codes. We perform these computations using MAGMA software \cite{bosma1997magma}.

 \section{$q$-ary LCD codes from $\mathbb{F}_{q}\mathbb{F}_{q^2}$-additive codes}\label{Sec5}
 The following result derives the $q$-ary LCD codes from $\mathbb{F}_{q}\mathbb{F}_{q^2}$-additive codes under certain assumptions. 
  \begin{theorem} \label{LCD theorem}
    Consider the $\mathbb{F}_{q}\mathbb{F}_{q^2}$-additive code $\mathcal{C}$, generated by the matrix $G = (G_{\alpha} \mid G_{\beta})$, where $G_{\alpha} \in M_{k\times\alpha}(\mathbb{F}_{q})$ and $G_{\beta} \in M_{k\times\beta}(\mathbb{F}_{q^2})$ and the code $\mathcal{C}_{\alpha} = \langle G_{\alpha} \rangle$ is self-orthogonal. Suppose $G_{\beta}$ is a matrix with rows linearly independent over $\mathbb{F}_{q}.$ If  $\phi(\mathcal{C}_{\beta})$ is LCD over $\mathbb{F}_{q}$, then so is $\Phi(\mathcal{C})$.
\end{theorem}

\begin{proof}
     Suppose $\Phi(\mathbf{u}\mid \mathbf{u}') \in \Phi(\mathcal{C})\cap \Phi(\mathcal{C})^{\perp}$ for some non-zero element $(\mathbf{u}\mid \mathbf{u}') \in \mathcal{C}$. Let $(\mathbf{v}\mid \mathbf{v}')$ be any arbitrary element of $\mathcal{C}$. Since $\mathcal{C}_{\alpha}$ is self-orthogonal, $[\mathbf{u},\mathbf{v}] = 0$, where $[\cdot,\cdot]$ is the standard inner product. Then
    \begin{eqnarray*}
          0 &=& [\Phi(\mathbf{u}\mid\mathbf{u}'), \Phi(\mathbf{v}\mid\mathbf{v}')] \\
          &=& [\mathbf{u},\mathbf{v}] + [\phi(\mathbf{u}'),\phi(\mathbf{v}')] \\
          &=& [\phi(\mathbf{u}'),\phi(\mathbf{v}')].
    \end{eqnarray*}
    Thus $\phi(\mathbf{u}') \in \phi(\mathcal{C}_{\beta}) \cap \phi(\mathcal{C}_{\beta})^{\perp} = \{\mathbf{0}\}$, which implies $\mathbf{u}' = \mathbf{0}$ as $\phi$ is injective. Suppose for $1\leq j \leq n$, $(\mathbf{y}_{j}\mid  \mathbf{y}_{j}')$ is the $j$-th row of $G$. Since $(\mathbf{u},\mathbf{u}') \in \mathcal{C}$, there exists $\mu_{1}, \mu_{2},\ldots, \mu_{k} \in \mathbb{F}_{q}$ such that $(\mathbf{u}|\mathbf{0}) = \sum\limits_{j = 1}^{k}\mu_{j}(\mathbf{y}_{j}\mid \mathbf{y}_{j}') = \left(\sum\limits_{j = 1}^{k} \mu_{j}\mathbf{y}_{j}\mid \sum\limits_{j = 1}^{k} \mu_{j}\mathbf{y}_{j}' \right)$. Since the set $\{\mathbf{y}_{1}',\mathbf{y}_{2}',\ldots,\mathbf{y}_{k}'\}$ is linearly independent over $\mathbb{F}_{q}$, $\mu_{j} = 0$ for all $j=1,2,\ldots,k$. Hence, $(\mathbf{u}\mid \mathbf{u}') = (\mathbf{0}\mid \mathbf{0})$, which is a contradiction.  
    \qed
\end{proof}
\begin{example}
    Let $q=3$ and let $\omega $ be a root of $x^2+1 \in \mathbb{F}_{3}[x]$. Then $\mathbb{F}_{9} = \mathbb{F}_{3}[\omega]$. Suppose $\mathcal{C}$ is a $\mathbb{F}_{3}\mathbb{F}_{9}$-additive code of length $4$ generated by the matrix 
    $$ G = \left(
    \begin{tabular}{c c c c|c c c c}
        $1$ & $1$ & $1$ & $0$ & $\omega $ & $\omega +1$ & $\omega +1$ & $\omega $\\
        $1$ & $2$ & $0$ & $1$ & $\omega +2$ & $2$ & $\omega $ & $1$\\
        $1$ & $2$ & $0$ & $1$ & $2$ & $\omega $ & $2$ & $\omega $
    \end{tabular}
    \right).$$
    One can verify that $\mathcal{C}_{\alpha}$ is self-orthogonal and $G_{\beta}$ has $\mathbb{F}_{3}$-linearly independent rows. Also, $\phi(\mathcal{C}_{\beta})$ is a $[8,3,4]$-LCD code over $\mathbb{F}_{3}$ with generator matrix 
    $$ \left(
\begin{tabular}{c c c c c c c c}
$1$ & $0$ & $0$ & $1$ & $2$ & $2$ & $2$ & $2$ \\
$0$ & $1$ & $0$ & $1$ & $0$ & $2$ & $0$ & $2$ \\
$0$ & $0$ & $1$ & $2$ & $1$ & $2$ & $1$ & $2$
\end{tabular}
\right).$$
Moreover, $\Phi(\mathcal{C})$ is a $[12,3,7]$-LCD code over $\mathbb{F}_{3}$ and a generator matrix is given by
\[ \left(
\begin{tabular}{c c c c c c c c c c c c }
$1$ & $0$ & $2$ & $2$ & $0$ & $0$ & $2$ & $1$ & $2$ & $1$ & $2$ & $1$ \\
$0$ & $1$ & $2$ & $1$ & $0$ & $1$ & $1$ & $0$ & $1$ & $1$ & $1$ & $1$ \\
$0$ & $0$ & $0$ & $0$ & $1$ & $1$ & $2$ & $0$ & $1$ & $2$ & $1$ & $2$
\end{tabular}
\right).
\]
Observe that $\Phi(\mathcal{C})$ is optimal according to \cite{makoto2021lcd}
\end{example}\par

Using the Gray map $\Phi$ in Eq. \eqref{phi}  and Theorem \ref{LCD theorem}, we find the LCD codes over $\mathbb{F}_{3}$ and the optimality of these codes is verified using \cite{makoto2021lcd} and \cite{pang2020bounds}. These codes are listed in Table \ref{table3}, where $\omega$ is a zero of the polynomial $x^2+1 \in \mathbb{F}_{3}[x]$. We perform all computations using MAGMA software \cite{bosma1997magma}.

\begin{landscape}
    \begin{table}
     \centering
     \scalebox{0.85}{
    \begin{tabular}{|l|l|p{4cm}|l|l|p{4cm}|l|l|l|}
    \hline
      \multirow{2}{*}{S. N.} & \multicolumn{5}{|c|}{Generators}  & \multirow{2}{*}{$[\alpha, \beta]$} & \multirow{2}{*}{Parameters} & \multirow{2}{*}{Remark}\\
\cline{2-6}
& $s(x)$ & $g(x)$ & $h(x)$ & $k(x)$ & $l(x)$ & & &\\
        \Xhline{3\arrayrulewidth}
   
   $1$ & $1$ & $ x+2$ & $ x+2$ & $x^3+2$ & $x^2+x+1$ & $[1, 3]$ & $[7, 3, 4]$ & Optimal \\
   
   $2$ & $1$ & $ 1$ & $ x^2+2x+2$ & $x^5+2$ & $x^2+(2\omega +2)x+\omega +1$ & $[1, 5]$ & $[11, 6, 5]$ & Optimal \\
   
   $3$ & $1$ & $ x^6+x^5+x^4+x^3+x^2+x+1$ & $ x^7+2$ & $x^7+2$ & $(2\omega +2)x^4+(\omega +1)y^3+2\omega y^2+(\omega +2)y+\omega +1$ & $[1, 7]$ & $[15, 8, 5]^{b,\ddagger}$ & Optimal \\

   $4$ & $1$ & $ x^8+x^7+x^6+x^5+x^4+x^3+x^2+x+1$ & $ x^9+2$ & $x^9+2$ & $x^4+(2\omega +1)x^3+(2\omega +1)x^2+(\omega +2)x+\omega $ & $[1, 9]$ & $[19, 9, 7]^{b,\ddagger}$ & Optimal \\  

   $5$ & $1$ & $ x^8+x^7+x^6+x^5+x^4+x^3+x^2+x+1$ & $ x^9+2$ & $x^9+2$ & $x^4+(2\omega +1)x^3+(2\omega +1)x^2+(\omega +2)x+2\omega +2$ & $[1, 9]$ & $[19, 10, 6]^{\ddagger}$ & Optimal \\ 
   
   $6$ & $1$ & $x^{13}+x^{12}+x^{11}+x^{10}+x^9+ x^8+x^7+x^6+x^5+x^4+x^3+x^2+x+1$ & $ x^{14}+2$ & $x^{14}+2$ & $(\omega +1)x^4+2\omega x^2+x+\omega $ & $[1, 14]$ & $[29, 15, 8]^{\ddagger}$ & BKLC \\
   
   $7$ & $1$ & $x^{16}+x^{15}+x^{14}+x^{13}+x^{12}+x^{11}+x^{10}+x^9+x^8+x^7+x^6+x^5+x^4+x^3+x^2+x+1$ & $ x^{17}+2$ & $x^{17}+2$ & $\omega x^8+(2\omega +2)x^7+\omega x^6+(\omega +1)x^5+(\omega +2)x^3+(\omega +2)x^2+(\omega +2)x+2\omega $ & $[1, 17]$ & $[35, 18, 11]$ & Optimal \\

   $8$ & $x^2+x+1$ & $ x^3+2$ & $ 2x^2+2x+2$ & $x^3+2$ & $(\omega +1)x^2+(\omega +1)x+(\omega +1)$ & $[3, 3]$ & $[9, 2, 6]^{a,\ddagger}$ & Optimal \\

   $9$ & $1$ & $ 1$ & $ x$ & $x^3+2$ & $2\omega +2$ & $[3, 3]$ & $[9, 6, 3]^{a,b}$ & Optimal \\

   $10$ & $x+2$ & $1$ & $x$ & $x^3+2x^2+x+2$ & $x+2\omega $ & $[3, 4]$ & $[11, 7, 3]^{\ddagger}$ & Optimal \\

   $11$ & $1$ & $ 1$ & $x^3+x^2+2x$ & $x^3+2x^2+x+2$ & $(2\omega +1)x^3+2x$ & $[3, 4]$ & $[11, 10, 2]^{b}$ & MDS \\

   $12$ & $x^2+x+1$ & $ x^6+x^5+x^4+x^3+x^2+x+1$ & $ x^7+2$ & $x^7+2$ & $(\omega +2)x^5+x^4+(2\omega +2)x^3+\omega x^2+2$ & $[3, 7]$ & $[17, 8, 6]^{b,\ddagger}$ & Optimal \\

   $13$ & $1$ & $ 1$ & $ x$ & $x^4+2$ & $2\omega +2$ & $[4, 4]$ & $[12, 8, 3]^{a,b,\ddagger}$ & Optimal \\

   $14$ & $1$ & $ 1$ & $x$ & $x+2$ & $(2\omega +2)x^3+\omega x^2+(2\omega +1)x)$ & $[4, 4]$ & $[12, 11, 2]^{a}$ & MDS \\
   
   \hline
    \end{tabular}}
            \begin{tablenotes}
            \item \textsuperscript{a}Quasi-cyclic code
            \item \textsuperscript{b}LCD code
            \item \textsuperscript{$\ddagger$}  codes that are not equivalent to the best-known linear codes
        \end{tablenotes}
    \caption{Gray image of $\mathbb{F}_{3}\mathbb{F}_{9}$-additive cyclic codes}
    \label{table2}
\end{table}
\end{landscape}
\begin{landscape}
\begin{table}[t]
    \centering
    \begin{tabular}{|c|c|c|c|c|c|}
\hline
    $q$ & $\alpha$ & $\beta$ & $G$ & Parameters for $\Phi(\mathcal{C})$ & Remarks  \\
\Xhline{3\arrayrulewidth}
$3$ & $4$ & $2$ & $ \left(
    \begin{tabular}{c c c c|c c }
         $1$ & $1$ & $1$ & $0$ & $\omega $ & $\omega $   \\
         $1$ & $2$ & $0$ & $1$ & $2$ & $\omega +1$  
    \end{tabular}
    \right) $ & $[8,2,5]$ & Optimal \\
    \hline
$3$ & $4$ & $2$ & $ \left(
    \begin{tabular}{c c c c|c c }
         $1$ & $1$ & $1$ & $0$ & $\omega $ & $\omega +1$   \\
         $1$ & $2$ & $0$ & $1$ & $\omega +2$ & $2$  \\
         $1$ & $2$ & $0$ & $1$ & $2$ & $\omega $   
    \end{tabular}
    \right) $ & $[8,3,4]$ & Optimal LCD code \\
    \hline
$3$ & $4$ & $3$ & $ \left(
    \begin{tabular}{c c c c|c c c}
         $1$ & $1$ & $1$ & $0$ & $\omega $ & $\omega $ & $1$   \\
         $1$ & $2$ & $0$ & $1$ & $2$ & $\omega +1$ & $\omega $  
    \end{tabular}
    \right) $ & $[10,2,7]$ & Optimal \\
    \hline
$3$ & $4$ & $3$ & $ \left(
    \begin{tabular}{c c c c|c c c}
         $0$ & $0$ & $0$ & $0$ & $\omega $ & $\omega +1$ & $\omega +1$   \\
         $1$ & $1$ & $1$ & $0$ & $\omega +2$ & $2$ & $\omega $  \\
         $1$ & $2$ & $0$ & $1$ & $2$ & $\omega +1$ & $2\omega $  
    \end{tabular}
    \right) $ & $[10,3,6]$ & Optimal \\
    \hline
$3$ & $4$ & $4$ & $ \left(
    \begin{tabular}{c c c c|c c c c}
         $1$ & $1$ & $1$ & $0$ & $\omega $ & $\omega $ & $1$ & $\omega $   \\
         $1$ & $2$ & $0$ & $1$ & $2$ & $\omega +1$ & $\omega $ & $1$  
    \end{tabular}
    \right) $ & $[12,2,8]$ & Optimal LCD code\\
    \hline
$3$ & $4$ & $4$ & $ \left(
    \begin{tabular}{c c c c|c c c c}
         $1$ & $1$ & $1$ & $0$ & $\omega $ & $\omega +1$ & $\omega +1$ & $\omega $   \\
         $1$ & $2$ & $0$ & $1$ & $\omega +2$ & $2$ & $\omega $ & $1$  \\
         $1$ & $2$ & $0$ & $1$ & $2$ & $\omega $ & $2$ & $\omega $
    \end{tabular}
    \right) $ & $[12,3,7]$ & Optimal LCD code\\
    \hline
$3$ & $4$ & $5$ & $ \left(
    \begin{tabular}{c c c c|c c c c c}
         $1$ & $1$ & $1$ & $0$ & $\omega $ & $\omega $ & $1$ & $\omega +2$ & $\omega $   \\
         $1$ & $2$ & $0$ & $1$ & $2$ & $\omega +1$ & $\omega $ & $1$ & $2$  
    \end{tabular}
    \right) $ & $[14,2,10]$ & Optimal \\
    \hline
$3$ & $4$ & $6$ & $ \left(
    \begin{tabular}{c c c c|c c c c c c}
         $1$ & $1$ & $1$ & $0$ & $\omega $ & $\omega $ & $1$ & $\omega +2$ & $\omega $ & $\omega $   \\
         $1$ & $2$ & $0$ & $1$ & $2$ & $\omega +1$ & $\omega $ & $1$ & $2$ & $\omega $  
    \end{tabular}
    \right) $ & $[16,2,11]$ & Optimal LCD code \\
    \hline
$3$ & $4$ & $6$ & $ \left(
    \begin{tabular}{c c c c|c c c c c c}
         $1$ & $1$ & $1$ & $0$ & $2\omega $ & $2\omega +2$ & $0$ & $\omega +1$ & $2\omega +2$ & $2\omega +1$   \\
         $1$ & $1$ & $1$ & $0$ & $2$ & $\omega +1$ & $2\omega $ & $0$ & $2$ & $\omega +1$ \\
         $1$ & $2$ & $0$ & $1$ & $2\omega +1$ & $0$ & $\omega +1$ & $2\omega +2$ & $1$ & $\omega +2$
    \end{tabular}
    \right) $ & $[16,3,10]$ & Optimal  \\
    \hline
$3$ & $4$ & $8$ & $ \left(
    \begin{tabular}{c c c c|c c c c c c c c}
         $1$ & $1$ & $1$ & $0$ & $\omega $ & $\omega $ & $1$ & $\omega +2$ & $\omega $ & $\omega +1$ & $2\omega $ & $2\omega +1$   \\
         $1$ & $2$ & $0$ & $1$ & $2$ & $\omega +1$ & $\omega $ & $1$ & $2$ & $\omega $ & $2$ & $2$  
    \end{tabular}
    \right) $ & $[20,2,14]$ & Optimal LCD code \\
\hline
\end{tabular}
    \caption{Ternary LCD codes from $\mathbb{F}_{3}\mathbb{F}_{9}$-additive codes}
    \label{table3}
\end{table}
\end{landscape}

\section{Conclusion}\label{Sec6}
In this article, we have studied  $\mathbb{F}_{q}\mathbb{F}_{q^2}$-additive cyclic codes. We have determined their generator polynomials and minimal spanning sets. Various examples of additive codes that attain the Singleton bound are displayed in Table \ref{table1}. By introducing a Gray map, we have obtained optimal linear codes over $\mathbb{F}_{3}$. Some of them are LCD and quasi-cyclic as well. These linear codes are listed in Table \ref{table2}. Finally, we establish a condition under which $\mathbb{F}_{q}\mathbb{F}_{q^2}$-additive codes are LCD over $\mathbb{F}_{q}$. Based on this result, we provide several optimal ternary LCD codes, which are shown in Table \ref{table3}.

In future, it may be worthwhile to investigate whether $\mathbb{F}_{q}\mathbb{F}_{q^2}$-additive cyclic codes are asymptotically good. One can also study  $\mathbb{F}_{q}\mathbb{F}_{q^t}$-additive cyclic codes for $t\geq 3$.

\subsection*{Acknowledgements}
The first author acknowledges MHRD, India, for financial support via a Junior Research Fellowship at IIT Delhi. The authors also express their gratitude to Dr. Gyanendra Kumar Verma for helpful discussions. The FIST Lab, reference number SR/FST/MS1/2019/45, was used for computational work.
\subsection*{Author Contributions}
Ankit Yadav and Ritumoni Sarma contributed equally to this work. Both authors read and approved the final manuscript.
\subsection*{Funding}
Not applicable.
\subsection*{Availability for supporting data} Not applicable.
\section*{Declarations}
\subsection*{Ethics approval and consent to participate}
Not applicable.
\subsection*{Consent for Publication} 
Not applicable.
\subsection*{Conflict of Interest}
Both authors declare that they have no conflicts of interest.

\bibliographystyle{abbrv}
\bibliography{ref}
\end{document}